\documentclass[showpacs,aps,prc,twocolumn,superscriptaddress,floatfix,nofootinbib,10pt]{revtex4-1}
\usepackage{amsmath,graphicx}
\usepackage[pdftitle={The Skyrme Tensor Force in Heavy Ion Collisions},pdfauthor={P. D. Stevenson},colorlinks=true,citecolor=black,linkcolor=black]{hyperref}
\usepackage{color} 

\begin{document}
\title{The Skyrme Tensor Force in Heavy Ion Collisions}
\author{P.~D.~Stevenson}
\affiliation{Department of Physics, University of Surrey, Guildford, Surrey, GU2 7XH, United Kingdom}
\author{E.~B.~Suckling}
\affiliation{Department of Physics, University of Surrey, Guildford, Surrey, GU2 7XH, United Kingdom}
\affiliation{Dept. of Meteorology, University of Reading, Reading, RG6 6BB, United Kingdom}
\author{S.~Fracasso}
\affiliation{Department of Physics, University of Surrey, Guildford, Surrey, GU2 7XH, United Kingdom}
\author{M.~C.~Barton}
\affiliation{Department of Physics, University of Surrey, Guildford, Surrey, GU2 7XH, United Kingdom}
\author{A.~S.~Umar}
\affiliation{Department of Physics and Astronomy, Vanderbilt University, Nashville, TN 37235, USA}
\date{\today}
\begin{abstract}
\begin{description}
\item[Background]
It is generally acknowledged that the time-dependent Hartree-Fock (TDHF) method provides a  useful foundation for a fully microscopic many-body theory of low-energy heavy-ion reactions. 
The TDHF method is most widely known in nuclear physics in the small amplitude domain, where it provides a useful description of collective states, and is based on the mean-field formalism which has been a relatively successful approximation to the nuclear many-body problem. Currently, the TDHF theory is  being widely used in the study of fusion excitation functions, fission, deep-inelastic scattering of heavy mass systems, while providing a natural foundation for many other studies.

\item[Purpose]
With the advancement of computational power it is now possible to undertake TDHF calculations without any symmetry assumptions and incorporate the major strides made by the nuclear structure community in improving the energy density functionals used in these calculations. In particular, time-odd and tensor terms in these functionals are naturally present during the dynamical evolution, while being absent or minimally important for most static calculations. The parameters of these terms are determined by the requirement of Galilean invariance or local gauge invariance but their significance for the reaction dynamics have not been fully studied. This work addresses this question with emphasis on the tensor force.

\item[Method]
The full version of the Skyrme force, including terms arising only from the Skyrme tensor force, is applied to the study of collisions within a completely symmetry-unrestricted TDHF implementation.

\item[Results]
We examine the effect on fusion thresholds with and without the tensor
force terms and find an effect on the fusion threshold energy of the
order several MeV.  Details of the distribution of the energy within
terms in the energy density functional is also discussed.

\item[Conclusions] 
Terms in the energy density functional linked to the tensor force can
play a non-negligible role in dynamic processes in nuclei, including
those in which the terms do not affect ground state properties.
\end{description}
\end{abstract}
\pacs{21.60.Jz}
\maketitle
\section{\label{sec:intro}Introduction}
The extent to which the time-dependent mean field is a sufficiently
good description of certain phenomena in nuclear dynamics is still an
open question, as evinced partly by the ongoing explorations and
developments in the
area~\cite{Si12,UO10a,SY13,II13,SK13,US14,SU14,UO06,Ste10,KU12,Eba10,Iwa10,Sim11,DaiPRC}. In
particular, calculations of nuclear fusion in a time-dependent
Hartree-Fock framework have developed over the years with increasingly
relaxed symmetry assumptions, and more sophisticated energy
functionals in an attempt to settle this question.  Approximations of
any type limit the number of degrees of freedom accessible during a
collision, and hence affect the nature and degree of dissipation. The
understanding of the dissipative mechanisms in the TDHF theory is
vital for establishing the region of validity of the mean-field
approximation and providing estimates for the importance of the
mean-field effects at higher energies. In TDHF, the dissipation of the
translational kinetic energy of the two ions is due to the collisions
of single particle states with the walls of the time-dependent
potential. This leads to the randomization of the motion characterized
by the distribution of energy among all possible degrees of freedom of
the system. The complete equilibration of the translational kinetic
energy among all possible degrees of freedom is commonly accepted as
being the definition of fusion whereas the incomplete equilibration
results in inelastic collisions. 

The inclusion of the spin-orbit interaction had a dramatic effect on
dissipation modes in heavy-ion collisions~\cite{RU89,US89}.  The
relaxation of any spatial symmetries gave rise to yet new
modes~\cite{Mar06}.  The inclusion of all time-odd terms that arise
from basic (not including tensor terms) Skyrme functionals was
analyzed and found to have a noticeable effect on fusion properties
\cite{Uma06} and in giant resonances~\cite{Nak05}.  These kinds of
calculations help to pin down details of the nuclear energy density
functional in ways complementary to studies of nuclear
matter~\cite{Rik03,Uma07,Dut12,Pas12} and of the structure of finite
nuclear structure~\cite{Les07,Suc10,Xia12}.  

In this work, we follow up previous studies by
analyzing the effect of using Skyrme functionals which include the tensor
part of the original Skyrme force, and we include the most general terms
in the density functional that arise, noting the interesting effects
that have been seen in calculations of giant resonances when adding the
tensor terms \cite{FraGDR}.

In Section~\ref{sec:skyrme} we briefly outline the Skyrme force and the resulting energy density functional used in our
analysis. Section~\ref{sec:analysis} presents TDHF calculations along
with a discussion of the results, with a focus on the tensor terms of
the functional. For comparison with previous work we concentrate on
$^{16}$O on $^{16}$O collisions, where the tensor force has little
effect on the static ground state properties, so that in addition to
comparison to previous studies, dynamic effects can be isolated.  A
concluding section (\ref{sec:conclusions}) follows the results.

\section{\label{sec:skyrme}The Skyrme Force and the Energy Density Functional}
Skyrme's interaction was originally posited as a zero-range low-momentum expansion of the effective interaction
in nuclear medium~\cite{Sky59}. The original form for the two-body part of the potential was given as $t_{12}=\delta(\textbf{r}_1-{\bf
r}_2)t({ \bf k}',\textbf{k})$, where
\begin{eqnarray}
t(\textbf{k}',\textbf{k}) &=& t_0(1+x_0P^\sigma)+\frac{1}{2}t_1(1+x_1P^\sigma)(\textbf{k}'^2+\textbf{k}^2)\nonumber\\
&+&t_2(1+x_2(P^\sigma-\frac{4}{5}))\boldsymbol{k}'\cdot\boldsymbol{k} \nonumber \\
&+&\frac{1}{2}T(\boldsymbol{\sigma}_1\cdot\boldsymbol{k}\boldsymbol{\sigma}_2\cdot\boldsymbol{k}-\frac{1}{3}\boldsymbol{\sigma}_1\cdot\boldsymbol{\sigma}_2\boldsymbol{k}^2 + \mathrm{conj.})) \nonumber \\
&+&\frac{1}{2}U(\boldsymbol{\sigma}_1\cdot\boldsymbol{k}'\boldsymbol{\sigma}_2\cdot\boldsymbol{k}-\frac{1}{3}\boldsymbol{\sigma}_1\cdot\boldsymbol{\sigma}_2\boldsymbol{k}'\cdot\boldsymbol{k}+\mathrm{conj.}) \nonumber \\
&+& V(i(\boldsymbol{\sigma}_1+\boldsymbol{\sigma}_2)\cdot\boldsymbol{k}'\times\boldsymbol{k}).
\label{eq:tkk}
\end{eqnarray}
In the above equation, the third and fourth lines represent the Skyrme tensor interaction, with $T$ and $U$ being parameters
to fit to data.  In the first Hartree-Fock calculations using the Skyrme interaction, in 1972, the tensor terms were
neglected~\cite{Vau72}.  This was reasonable, since only ground states of doubly-magic nuclei were calculated.  The
contribution of the tensor terms is, in that case, mainly to the spin-orbit splitting, which already has an adjustable
parameter ($V$ in (\ref{eq:tkk})), and the data to which the force was
fitted did not demand extra parameters in the spin-orbit part.  

The effect of the tensor terms was studied later, in 1977, again in doubly-magic nuclei~\cite{Sta77}, with mixed
conclusions as to their efficacy.  Sporadically, parameter sets including tensor terms were explored, culminating in a
recent resurgence in their study, motivated initially by the observed changing shell structure away from stability.  The
comprehensive paper by Lesinki et al., gives a summary of the history of the Skyrme tensor term, and we refer
the reader there for a more complete account~\cite{Les07}.

It is now routine to present the Skyrme interaction as an energy density functional (EDF).  This is a more physical
approach given that the original three-body force has been generalized to a density-dependent two-body term with a
fractional power of the density.  Following Lesinski et al.~\cite{Les07}, we present the full form of the Skyrme EDF,
including terms which arise from the tensor force:

\begin{eqnarray}
\mathcal{E} &=& \int d^3r\sum_{t=0,1}\Bigg\{C_t^\rho[\rho_0]\rho_t^2+C_t^s[\rho_0]\boldsymbol{s}_t^2+C_t^{\Delta\rho}\rho_t\nabla^2\rho_t\nonumber \\
&+&C_t^{\nabla s}(\nabla\cdot\boldsymbol{s})^2+C_t^{\Delta s}\boldsymbol{s}_t\cdot\nabla^2\boldsymbol{s}_t+C_t^{\tau}(\rho_t\tau_t-\boldsymbol{j}_t^2) \nonumber \\
&+&C_t^T\left(\boldsymbol{s}_t\cdot\boldsymbol{T}_t-\sum_{\mu,\nu=x}^zJ_{t,\mu\nu}J_{t,\mu\nu}\right) +C_t^F\Bigg[\boldsymbol{s}_t\cdot\boldsymbol{F}_t\nonumber\\
&&-\frac{1}{2}\left(\sum_{\mu=x}^zJ_{t,\mu\mu}\right)^2-\frac{1}{2}\sum_{\mu,\nu=x}^zJ_{t,\mu\nu}J_{t,\nu\mu}\Bigg]\nonumber\\
&+&C_t^{\nabla\cdot J}\left(\rho_t\nabla\cdot\boldsymbol{J}_t+\boldsymbol{s}_t\cdot\nabla\times\boldsymbol{j}_t\right)\Bigg\},
\label{eq:edens}
\end{eqnarray}
where the following densities and currents are defined~\cite{Les07,Dud00,Per04} in terms of the density matrix in
coordinate space for protons and neutrons (indicated by the subscript $q$):
\begin{equation}
\rho_q(\boldsymbol{r}\sigma,\boldsymbol{r}'\sigma')=\frac{1}{2}\rho_q(\boldsymbol{r},\boldsymbol{r}')\delta_{\sigma\sigma'}+\frac{1}{2}\boldsymbol{s}_q(\boldsymbol{r},\boldsymbol{r}')\cdot\langle\sigma'|\hat{\boldsymbol{\sigma}}|\sigma\rangle,
\end{equation}
with
\begin{equation}
\rho_q(\boldsymbol{r},\boldsymbol{r}')=\sum_\sigma\rho_q(\boldsymbol{r}\sigma,\boldsymbol{r}'\sigma')
\end{equation}
and
\begin{equation}
\boldsymbol{s}_q(\boldsymbol{r},\boldsymbol{r}')=\sum_{\sigma\sigma'}\rho_q(\boldsymbol{r}\sigma,\boldsymbol{r}'\sigma')\langle\sigma|\hat{\boldsymbol{\sigma}}|\sigma\rangle.
\end{equation}

The densities and currents needed are then defined in terms of the density
$\rho_q(\boldsymbol{r},\boldsymbol{r}')$ and spin density $\boldsymbol{s}_q(\boldsymbol{r},\boldsymbol{r}')$ as

\begin{eqnarray}
\rho_q(\boldsymbol{r})&=&\left.\rho_q(\boldsymbol{r},\boldsymbol{r}')\right|_{\boldsymbol{r}=\boldsymbol{r}'}\nonumber \\
\boldsymbol{s}_q(\boldsymbol{r}) &=& \left.\boldsymbol{s}_q(\boldsymbol{r},\boldsymbol{r}')\right|_{\boldsymbol{r}=\boldsymbol{r}'}\nonumber \\
\tau_q(\boldsymbol{r}) &=& \left.\nabla\cdot\nabla'\rho_q(\boldsymbol{r},\boldsymbol{r}')\right|_{\boldsymbol{r}=\boldsymbol{r}'} \nonumber \\
T_{q,\mu}(\boldsymbol{r}) &=& \left.\nabla\cdot\nabla's_{q,\mu}(\boldsymbol{r},\boldsymbol{r}')\right|_{\boldsymbol{r}=\boldsymbol{r}'}\\
\boldsymbol{j}_q(\boldsymbol{r}) &=& \left.-\frac{i}{2}(\nabla-\nabla')\rho_q(\boldsymbol{r},\boldsymbol{r}')\right|_{\boldsymbol{r}=\boldsymbol{r}'}\nonumber \\
J_{q,\mu\nu}(\boldsymbol{r})&=&\left.-\frac{i}{2}(\nabla_\mu-\nabla_\mu')s_{q,\nu}(\boldsymbol{r},\boldsymbol{r}')\right|_{\boldsymbol{r}=\boldsymbol{r}'}\nonumber\\
F_{q,\mu}(\boldsymbol{r})&=&\left.\frac{1}{2}\sum_{\nu=x}^z(\nabla_\mu\nabla_\nu'+\nabla_\mu'\nabla_\nu)s_{q,\nu}(\boldsymbol{r},\boldsymbol{r}')\right|_{\boldsymbol{r}=\boldsymbol{r}'}\nonumber,
\end{eqnarray}
where the Greek letter subscripts indicate Cartesian coordinates.  From these densities one then defines the isoscalar
($t=0$) and isovector ($t=1$) densities and currents found in (\ref{eq:edens}) as 
\begin{eqnarray}
\rho_0(\boldsymbol{r}) &=& \rho_n(\boldsymbol{r})+\rho_p(\boldsymbol{r}) \nonumber \\
\rho_1(\boldsymbol{r}) &=& \rho_n(\boldsymbol{r})-\rho_p(\boldsymbol{r}),
\end{eqnarray}
and similarly of the other densities and currents.

In the version of the energy density functional presented in
(\ref{eq:edens}), the contribution of the terms bilinear in the
spin-current pseudotensor $J_{\mu\nu}$ are presented in Cartesian form.
Some authors \cite{Per04} use a pseudoscalar, vector and rank-2
pseudotensor representation instead.  Denoting these as $J_t^{(0)}$,
$\mathbf{J}_t^{(0)}$, and $J_{t,\mu\nu}^{(0)}$ respectively, the
combinations of the $J^2$ pseudotensor that appear in the energy
density functional can alternatively be expressed as \cite{Les07}:
\begin{widetext}
\begin{eqnarray}
  \sum_{\mu,\nu=x}^z
  J_{t,\mu\nu}J_{t,\mu\nu}&=&\frac{1}{3}{J_t^{(0)}}^2+\frac{1}{2}\mathbf{J}_t^2+\sum_{\mu\nu=x}^zJ_{t,\mu\nu}^{(2)}J_{t,\mu\nu}^{(2)},\label{eq:coupled1}\\
  \frac{1}{2}\left(\sum_{\mu=x}^z
  J_{t,\mu\,mu}\right)^2+\frac{1}{2}\sum_{\mu,\nu=x}^z
  J_{t,\mu\nu}J_{t,\nu\mu} &=& \frac{2}{3}{J_t^{(0)}}^2-\frac{1}{4}\mathbf{J}_t^2+\frac{1}{2}\sum_{\mu\nu=x}^zJ_{t,\mu\nu}^{(2)}J_{t,\mu\nu}^{(2)}\label{eq:coupled2}
\end{eqnarray}
\end{widetext}

We compute and present these terms in both ways both as a check of the
implementation and to understand the ways in which the density
functional distributes energy across different terms.

We point out also, that in (\ref{eq:edens}), the tensor force parameters
$T$ and $U$ wholly determine coupling constants $C_t^{\nabla s}$ and
$C_t^F$, while contributing also, along with the central terms, to
$C_t^T$ and $C_t^{\Delta s}$.  Full details of the mapping between
Skyrme force parameterizations and energy density functional
coefficients are well-documented \cite{Les07}.  We take all terms in
(\ref{eq:edens}), as defined by this mapping, \textit{except} that we
set $C_t^{\nabla s}=C_t^{\Delta s}=0$ since these terms are known to
cause spin instabilities.  The inclusion of the tensor force  brings
new terms in the density functional in to play, as well as modifying
the strength of pre-existing terms. We note that a recent study
analyzed the effect of the tensor parameters on the spin-orbit force
alone (i.e. not including the ``new'' terms in the functional) in the
context of heavy-ion collisions \cite{Dai14}.  One does not need to
make a necessary link between the coefficients in a  force
representation to the coefficients of the EDF.  The latter can be
considered as the free parameters to be fitted to data.  It is not the
purpose of the present paper to produce fits, however, but rather to
evaluate the role of the terms that arise from the tensor force, and
the effect on non-zero tensor force parameters on other terms in the
EDF.  We use existing Skyrme+tensor fits from the literature,
motivated as follows. 

We choose SLy5 \cite{Cha98} as our test-bed, since there is a version
available which has had the tensor forces added perturbatively to give
improved single particle spectra in heavy isotopes without affecting
the bulk properties~\cite{Col07}.  It has also been extensively tested
in TDHF collision calculations in which the full $J^2$ tensor was
activated (as it was in the original SLy5 fit~\cite{Cha98}) and
time-odd terms studied~\cite{Uma06}.  Therefore, with this choice we
will not have to re-fit a force, and are able to compare with the
previous complete calculations of collisions in TDHF, albeit with no
tensor force. 

To test systematic properties of the tensor force parameters, we also examine the T$IJ$ forces~\cite{Les07}, in which the
isovector and isoscalar parts of the tensor force are systematically varied, though fitted only through their
contribution to the $J^2$ tensor.

\section{\label{sec:analysis}$^{16}$O + $^{16}$O Collisions}
Our calculations are performed on a Cartesian coordinate space grid with no symmetry assumptions, using a version of the
Sky3D code~\cite{sky3d} with all time-odd and tensor terms included.  In the first step, the ground state of $^{16}$O is
calculated with a damped gradient operator~\cite{BL92,BSU}. Iterations continue until the variance in the single particle
Hamiltonian
is sufficiently small that the time-dependent calculation will be stable and the nucleus will translate without loss of
energy on the grid~\cite{Mar06}.  We turn off any center of mass correction to be consistent between the static and
dynamic calculations.  Further details of the set-up of our method can be found
elsewhere \cite{EmmaThesis}

\subsection{Upper Fusion Thresholds for head-on collisions}
The fusion window for heavy-ion collisions occurs when the nuclei have sufficient kinetic energy to overcome the Coulomb
barrier, and sufficiently little that it can all be transferred to internal energy of the compound nucleus which then
stays fused.  Above the \textit{upper fusion threshold}, the collision is deep-inelastic in nature and no fusion occurs. 
The fusion threshold can be rather sensitive to the energy density functional, particularly when different terms come in
to play, which are not active in the ground states.  We repeat previous calculations for SkM*~\cite{Uma06}, as a check,
in which it is found that activating the $J^2$ terms (which are time-even) using the standard link between the Skyrme
parameters and those in (\ref{eq:edens}), a reduction in the fusion threshold of 6 MeV is found.  We reproduced the
size of the reduction, but found that activating the time-odd terms
increased slightly the fusion threshold, bringing it slightly closer
to the ``basic'' SkM* force,  as shown in
Table \ref{tab:fust}

\begin{table}[bht]
\centering
\begin{tabular}{cc}\hline
Force &  Threshold (MeV) \\ \hline\hline
SkM$^*$ (basic) & 77 \\ 
SkM$^*$ (inc. $J^2$) & 71  \\ 
SkM$^*$ (full) & 73  \\ 
SLy5 (full)  & 68  \\ 
SLy5t  & 65  \\ 
$T{12}$ & 61  \\ 
$T{14}$ & 69  \\ 
$T{22}$ & 64  \\ 
$T{24}$ & 71  \\ 
$T{26}$ & 82  \\ 
$T{42}$ & 69  \\ 
$T{44}$ & 79  \\ 
$T{46}$ & 87  \\ \hline\hline
\end{tabular}
\caption{Upper fusion threshold energies for the $^{16}$O + $^{16}$O collision using various parameterizations of the
Skyrme interaction.}
\label{tab:fust}
\end{table}
One sees that the tensor terms, when added to an existing
parameterizations (i.e. SLy5t compared to SLy5) can have a
non-negligible effect of around 5\% on the upper fusion threshold.
The variation in results from the T$IJ$ parameterizations is more
pronounced, yielding around a 25 MeV variation in the location of the
upper fusion threshold.  The lower threshold in each case is
determined by the Coulomb barrier, so is equal between the forces.
The fusion window, therefore, can be varied widely by the tensor part
of the Skyrme interaction. 

\subsection{EDF contributions in deep-inelastic collisions}
\begin{figure}[!hbt]
\includegraphics*[width=0.5\textwidth]{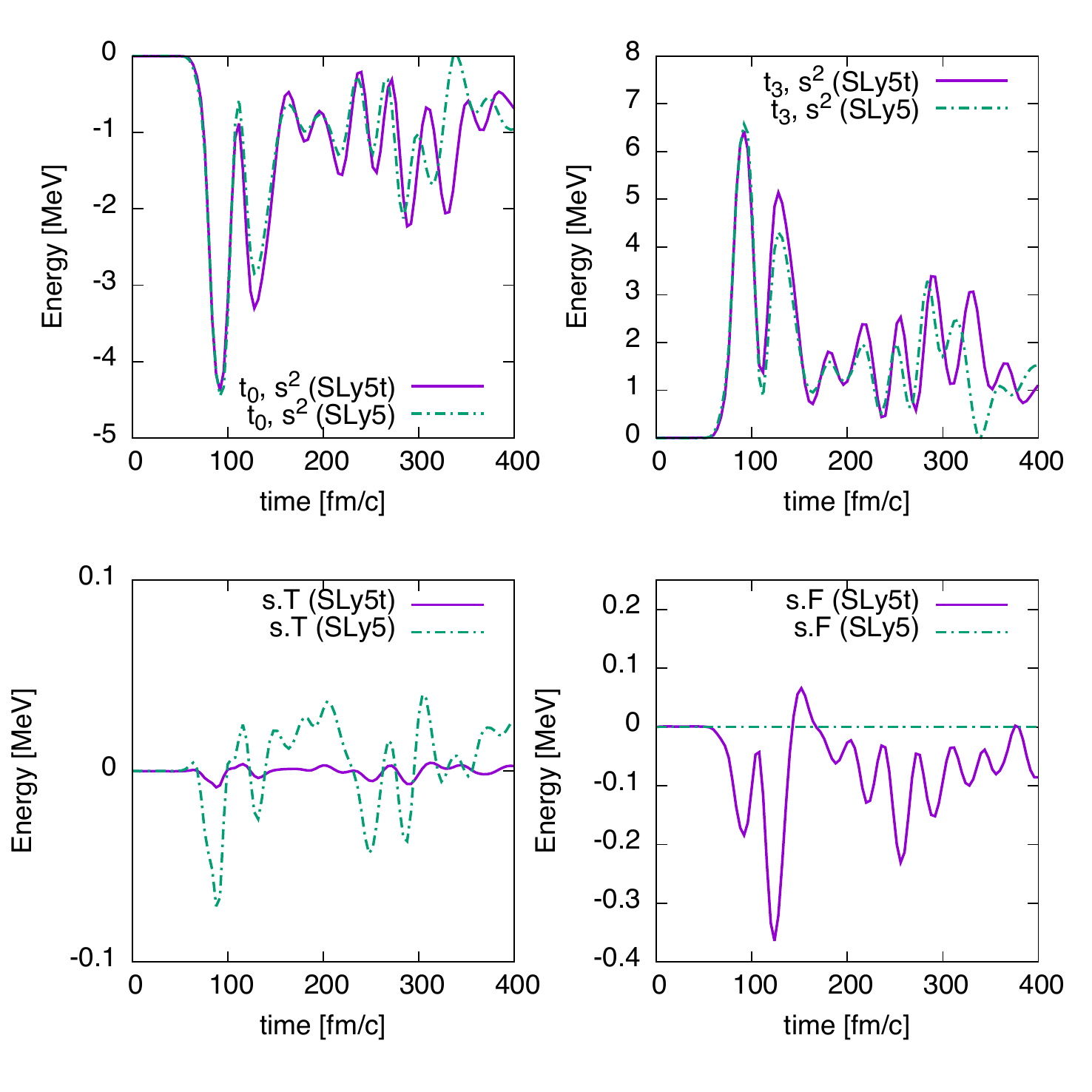}
\caption{(Color online) Contributions from terms involving some of the time-odd densities and currents to the total energy for 100 MeV collisions of $^{16}$O on $^{16}$O.\label{fig:contribs1}}
\end{figure}
We examine the contribution from different parts of the Skyrme Energy
Density Functional in the regime above the upper fusion threshold, so
that the collision separates into parts before collision, during it,
and afterwards when two nuclei re-emerge with internal excitation.
For $^{16}$O on $^{16}$O at 100 MeV center of mass energy we look at
SLy5 and SLy5t to examine the different contribution from the
different terms.  Figure \ref{fig:contribs1} shows the contribution
from the $s^2$ (split into those parts from the $t_0$ and $t_3$
parameters), the $\boldsymbol{s}.\boldsymbol{T}$ and the
$\boldsymbol{s}.\boldsymbol{F}$ terms.  Respectively, these are 
terms which do not depend directly on the tensor coefficients ($s^2$), are
amended compared to the non-tensor case
($\boldsymbol{s}.\boldsymbol{T}$) and
only come into play with non-zero tensor parameters ($\boldsymbol{s}.\boldsymbol{F}$).

As shown in the two upper panels of Figure \ref{fig:contribs1}, the
$s^2$ terms begin with zero contribution, as they should with two
$^{16}$O nuclei in their ground states, initialised with a Galilean
velocity boost.  Only when the nuclei begin to collide do these terms
begin to differ from zero, as the collision process gives rise to
regions of localised spin polarisation dynamically within the compound
nucleus.  Shortly after the collision, the $s^2$ contribution of the
SLy5 and SLy5t forces does not differ, as expected since the coupling
constants have not changed.  Only later when the presence of the
tensor terms has changed the overall dynamics do the details of the
$s^2$ contributions differ between the forces, owing to changes in the
$\boldsymbol{s}$ density itself. 

A much more evident change in the dynamics is seen in the
$\boldsymbol{s}.\boldsymbol{T}$ and $\boldsymbol{s}.\boldsymbol{F}$ terms.  Without the tensor
parameters, the $\boldsymbol{s}.\boldsymbol{F}$ term is identically
zero, while it is activated with a few hundred keV of the available
energy during collision via coupling to the tensor parameters.  The
$\boldsymbol{s}.\boldsymbol{T}$ terms, on the other 
hand, have couplings combining the tensor and surface terms of the
Skyrme interaction that conspire to much reduce the role of this term
when the tensor terms are activated in SLy5t, compared to SLy5.

Both the terms $\boldsymbol{s}.\boldsymbol{T}$ and
$\boldsymbol{s}.\boldsymbol{F}$ are linked by Galilean invariance to
terms arising from bilinear couplings of the spin-current tensor $J$,
as given by those terms in (\ref{eq:edens}) which share the same
parameter. 

\begin{figure}[!tbh]
\includegraphics*[width=0.5\textwidth]{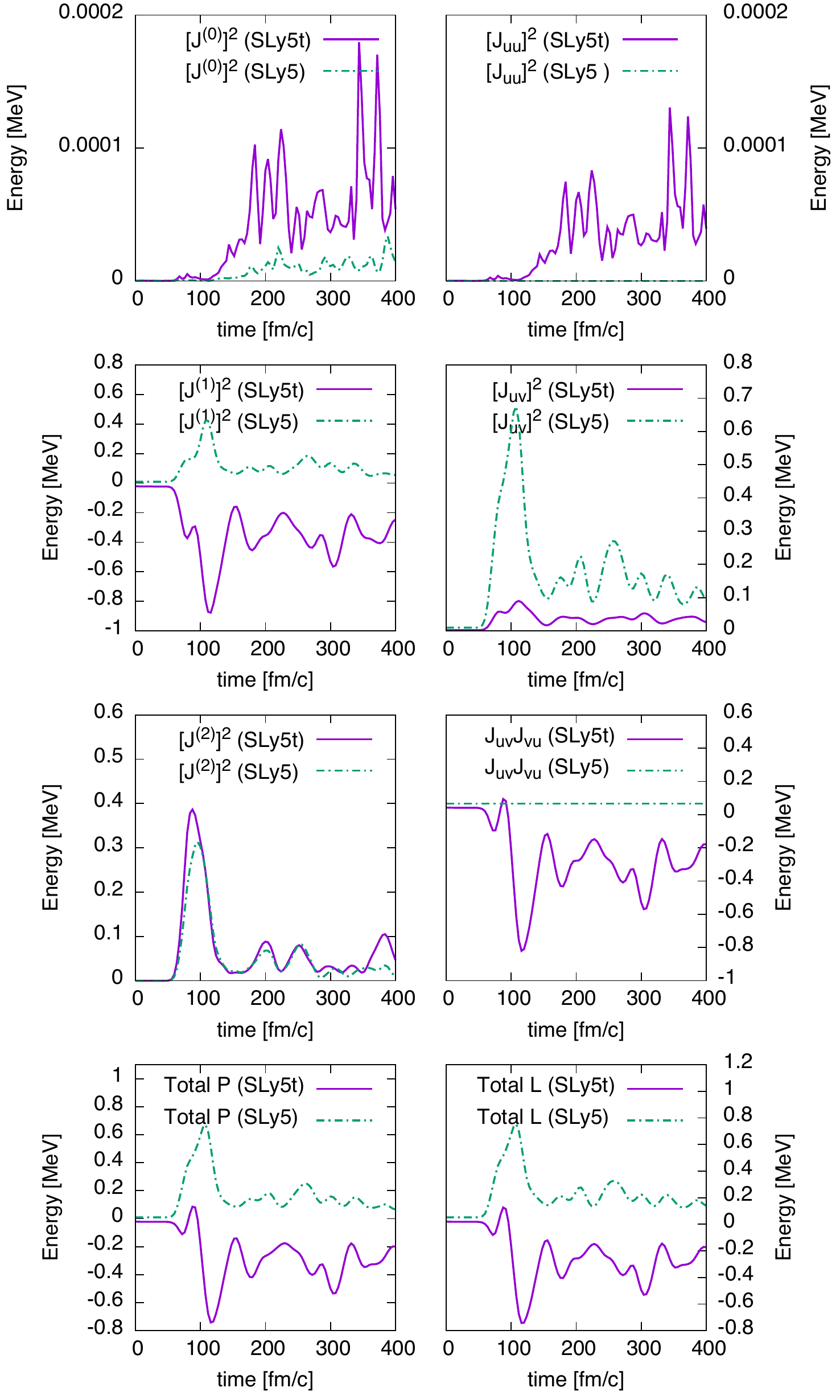}
\caption{(Color online) Contributions to the total energy from the $J^2$ terms
  following the Cartesian decomposition as in the paper by Lesinski
  \cite{Les07} (right column) or following the coupled form given by
  Perlinska \cite{Per04} (left column).  The bottom frames show the
  total contribution in both cases.  The case is 100 MeV collisions of $^{16}$O on $^{16}$O.\label{fig:jcontribs}}
\end{figure}

As a check of our approach, and also to understand the underlying
dynamics and interplay between the terms in the functional, we
evaluate the contributions due to the $J^2$ terms both in their
Cartesian form as given in (\ref{eq:edens}), and in their coupled form
given in (\ref{eq:coupled1}) and (\ref{eq:coupled2}).
Figure \ref{fig:jcontribs} shows these contributions.  The column on
the left shows the energy contribution from the (pseudo)scalar-, vector- and
(pseudo)tensor-decomposed form of $J^2$ while the right hand side shows those
as they appear in (\ref{eq:edens}).  The first three plots in the
right column are the diagonal, antisymmetric and symmetric terms
respectively.  As expected, the first and third terms are identically
zero in the case of SLy5 since they appear only multiplied by the
tensor coefficients.  The bottom frames of figure
{\ref{fig:jcontribs}} show the total contribution when calculated by
both approaches.  They are identical, as they must be.

\subsection{Off-axis collisions}
\begin{figure}[!hbt]
\includegraphics*[width=0.5\textwidth]{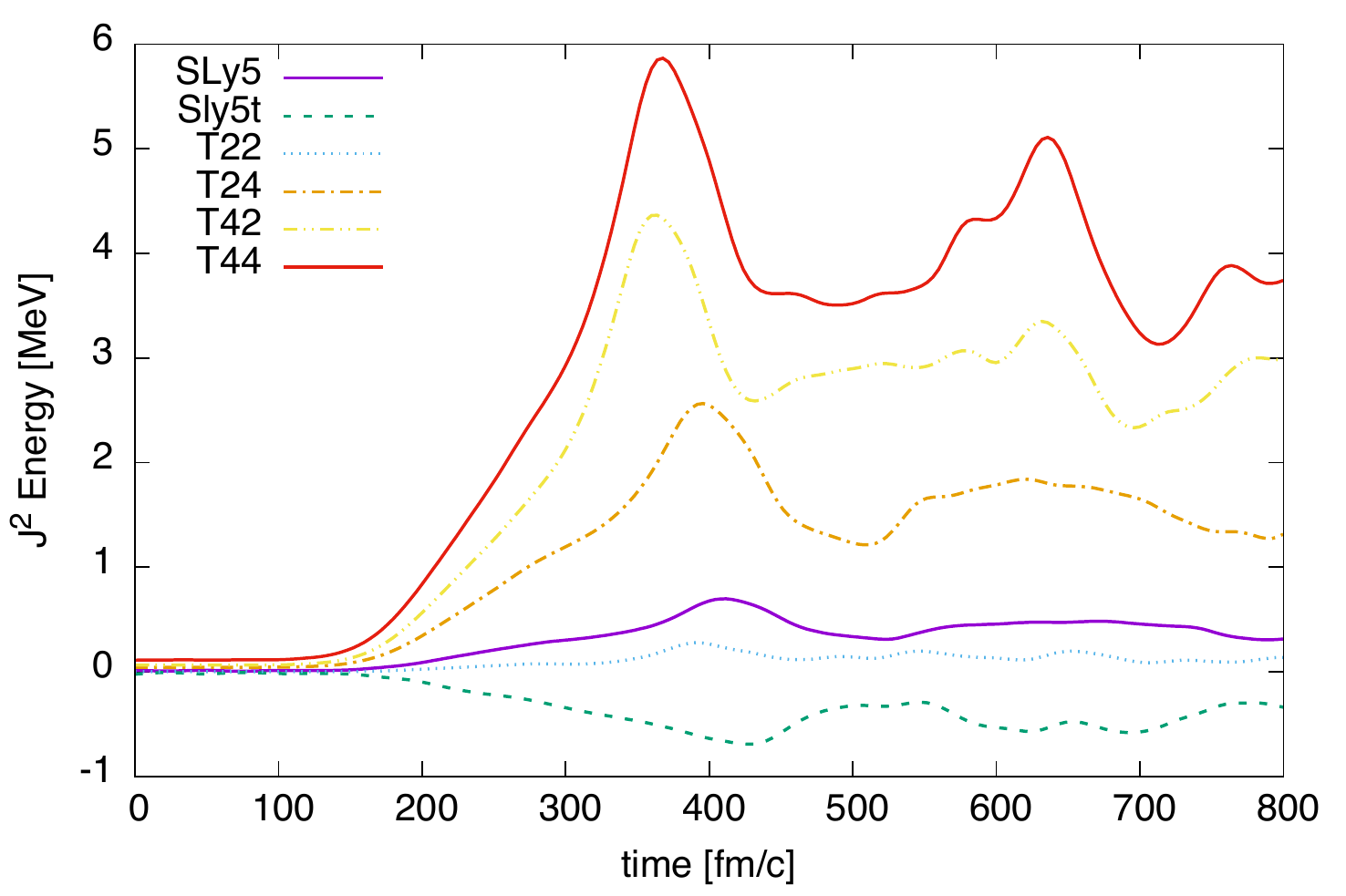}
\caption{(Color online) Contribution to the total energy from the $J^2$ terms for a $b=6.65$fm collision, which just fuses, at 34 MeV as a function of force.\label{fig:j-665}}
\end{figure}
For a series of different parameterisations, we performed calculations
of $^{16}$O on $^{16}$O at a center-of-mass energy of 34 MeV and
impact parameter close to a grazing impact, to validate against
previous work~\cite{Uma06} for the tensor-less forces and to
understand the dynamics in the nucleus.  In Figure \ref{fig:j-665} we
plot the contribution to the total energy of the system as described,
at $b=6.65$ fm, from the total $J^2$ terms as a function of the force
parameters.  It is seen that the perturbative addition of the tensor
terms to the SLy5t forces results in a different sign of the $J^2$
contribution to the mean field.  The results from the $TIJ$ forces
span a range of contribution, with up to several MeV of energy being
stored in this term at times during the collision process.  The $J^2$
terms are not identically zero at $t=0$, but they are greatly excited
during the collision process.  It is clear that the different
behaviour of different tensor parameterisations can appear in such a
dynamic situation, while  being much less evident in the ground state
(to which the force is fitted).

We mention in this section that the impact parameter dividing those
configurations which fuse from those which don't was rather
insensitive to the tensor parameter set, at least for this center of
mass energy of 34 MeV.  There is thus little effect on the
cross-section.   The study of Dai et al. \cite{Dai14} in which the tensor contribution to the spin-orbit interaction was evaluated, a variation in the cross-section is seen at 70.5 MeV.  This is consistent with the fact that we see a resonable spread of upper thresholds around this energy.

\section{Conclusions\label{sec:conclusions}}
We have performed time-dependent energy density functional
calculations of heavy ion collisions using $^{16}$O+$^{16}$O as a test
case.  We have included in the density functional all terms that arise
when deriving the functional from a Skyrme force including tensor
terms, and used existing Skyrme+tensor parameterisations to assess the
effect of the tensor terms within this framework.  It is found that
the size of the fusion window can vary as the tensor force varies,
owing to a movement in the upper fusion threshold.  Contributions from
different terms in the functionals were analysed.  Large variations in
the dominant $J^2$ terms, which could be either attractive or
repulsive, were found for different tensor-dependent functionals.

Terms in the density functional which arise from the tensor term,
which play a minor role in fitting ground state data, can have a
significant effect in dynamical properties.  Thus there is scope for
adjustment of functionals (or forces) in light of new data on
dynamical processes while retaining good fits to ground state data.

\begin{acknowledgments}
This work was funded by the UK Science and Technology Facilities
Council (STFC) under grant numbers ST/J00051/1, ST/J500768/1 and
ST/M503824/1 and by the U.S. Department of Energy under grant
No. DE-FG02-96ER40975 with Vanderbilt University.  We gratefully
acknowledge running time on the STFC DiRAC computer system.  One of us
(Stevenson) thanks Sigourney Fox for her hospitality during a visit to
Queen's University, Kingston, ON, during the preparation of this
manuscript.    
\end{acknowledgments}

\end{document}